\documentclass[10pt,article]{IEEEtran}

\usepackage{subcaption}
\usepackage{graphicx}
\usepackage{url}
\usepackage{amsmath}
\usepackage{amssymb}
\usepackage{color, colortbl}
\usepackage{import}
\usepackage{todonotes}
\usepackage{soul}

\definecolor{Gray}{gray}{0.8}
\definecolor{Lightgray}{gray}{0.9}
\graphicspath{{figures/}}

\begin{document}

\title{In-body Bionanosensor Localization for Anomaly Detection via Inertial Positioning and THz Backscattering Communication}

\author{
\IEEEauthorblockN{Jennifer Simonjan$^{*}$, Bige D. Unluturk$^{+}$, Ian F. Akyildiz$^{*}$}\\
\IEEEauthorblockA{$^{*}$Georgia Institute of Technology, GA, USA, \\
$^{+}$Michigan State University, MI, USA \\
Email: \{jennifer.simonjan@ieee.org\}\{unluturk@msu.edu\}\{ian@ece.gatech.edu\}}}

\maketitle

\begin{abstract}
Nanotechnology is enabling the development of a new generation of devices which are able to sense, process and communicate,
while being in the scale of tens to hundreds of cubic nanometers. Such small, imperceptible devices enhance not only current applications but enable entirely new paradigms especially for in-body environments. This paper introduces a localization and tracking concept for bionanosensors floating in the human bloodstream to detect anomalies in the body. Besides the nanoscale sensors, the proposed system also comprises macroscale anchor nodes attached to the skin of the monitored person. To realize autonomous localization and resource-efficient wireless communication between sensors and anchors, we propose to exploit inertial positioning and sub-terahertz backscattering. The proposed system is a first step towards early disease detection as it aims at localizing body regions which show anomalies. Simulations are conducted to enable a systematical evaluation on the feasibility of the approach.
\end{abstract}

\section{Introduction}

Novel nanomaterials such as graphene and its derivatives have made it possible to fabricate tiny sensors, i.e., bionanosensors (BNSs),
which are capable of detecting the smallest changes in physical variables, such as pressure, vibrations, temperature and concentrations in chemical and biological molecules \cite{panigrahi2018energy}.
BNSs enable to detect and sense molecules which are too small to be detected otherwise. One reason for late diagnosis of diseases such as cancer, is that current sensing technology can only detect biomarker molecules once they are manifested in the organs and have a sufficiently high molecule concentration. BNSs are however capable of detecting much lower concentrations and thus enable to detect diseases at a much earlier development stage \cite{singh2020nanosensors,munawar2019nanosensors}. As an example, glucose molecules and cancer cells have sizes in the range of a few nano- to micrometers. The only possibility to detect such small molecules is to have sensors with extraordinary high sensitivity while being in the size of the biomolecules themselves. BNSs are already in the development and will revolutionize in-body applications in the near future~\cite{akyildiz2015internet,akyildiz2020panacea}. 

To be able to detect diseased cells in the human body, hundreds of BNSs are foreseen to float through the cardiovascular system in order to check for anomalies. Whenever an anomaly occurs, the sensors should detect and report it to the outside world. A major concern in this field is thus to localize the body regions at which anomalies have been detected. Realizing such a localization system is very challenging due to the highly dynamic environment, the extremely constrained resources of the sensors and the difficulty of wireless communication through human tissue.

\begin{figure}[t]
	\centering
	\def\svgwidth{0.22\textwidth}
\begingroup%
  \makeatletter%
  \providecommand\color[2][]{%
    \errmessage{(Inkscape) Color is used for the text in Inkscape, but the package 'color.sty' is not loaded}%
    \renewcommand\color[2][]{}%
  }%
  \providecommand\transparent[1]{%
    \errmessage{(Inkscape) Transparency is used (non-zero) for the text in Inkscape, but the package 'transparent.sty' is not loaded}%
    \renewcommand\transparent[1]{}%
  }%
  \providecommand\rotatebox[2]{#2}%
  \newcommand*\fsize{\dimexpr\f@size pt\relax}%
  \newcommand*\lineheight[1]{\fontsize{\fsize}{#1\fsize}\selectfont}%
  \ifx\svgwidth\undefined%
    \setlength{\unitlength}{380.55191641bp}%
    \ifx\svgscale\undefined%
      \relax%
    \else%
      \setlength{\unitlength}{\unitlength * \real{\svgscale}}%
    \fi%
  \else%
    \setlength{\unitlength}{\svgwidth}%
  \fi%
  \global\let\svgwidth\undefined%
  \global\let\svgscale\undefined%
  \makeatother%
  \begin{picture}(1,1.60489258)%
    \lineheight{1}%
    \setlength\tabcolsep{0pt}%
    \put(0,0){\includegraphics[width=\unitlength,page=1]{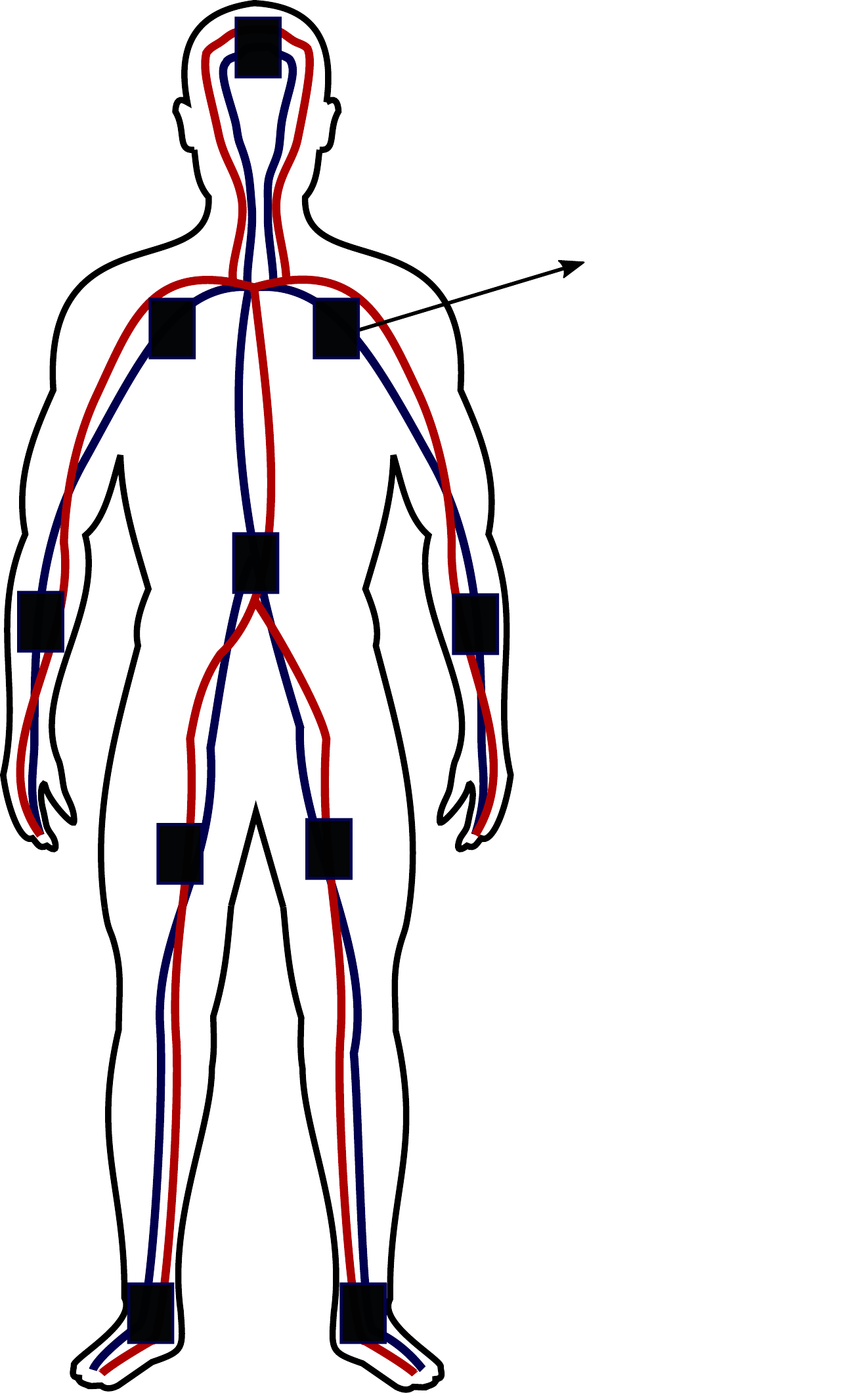}}%
    \put(0.78829647,1.30974481){\color[rgb]{0,0,0}\makebox(0,0)[lt]{\lineheight{1.25}\smash{\begin{tabular}[t]{l}Anchor\\nodes\end{tabular}}}}%
    \put(0,0){\includegraphics[width=\unitlength,page=2]{HumanContour.pdf}}%
    \put(0.86299705,0.39396902){\color[rgb]{0,0,0}\makebox(0,0)[lt]{\lineheight{1.25}\smash{\begin{tabular}[t]{l}Bionano-\\sensors\end{tabular}}}}%
    \put(0,0){\includegraphics[width=\unitlength,page=3]{HumanContour.pdf}}%
  \end{picture}%
\endgroup%

		\caption{Contour of a human body including the major vessels. Macro-scale anchor nodes are attached to the skin while bionanosensors float through the cardiovascular system.}
	\label{human}
\end{figure}

In this paper, we propose a system architecture to localize anomalies in the body detected by BNS measurements. Such a system is intended to be used for in-body health monitoring applications. Figure \ref{human} depicts an overview of the foreseen system which comprises anchor nodes and BNSs. Anchor nodes are of macroscale size and are attached to the skin of the monitored person while the sensors are of nanoscale size and float through the bloodstream. Due to the high mobility and limited communication range of the sensors, it is infeasible for the anchors to continuously locate and update the sensors. Hence, we require the sensors to track their location themselves and inform the anchors about the location of detected anomalies.
To do so, we propose to equip the BNSs with Inertial Measurement Units (IMUs) composed of nanoscale accelerometers and gyroscopes. Having access to IMU readings, sensors are able to exploit inertial positioning and can thus add location information to their actual sensor readings (i.e., temperature, vibration, molecule concentration). 

Furthermore, BNSs need to communicate their findings to the anchors in order to make them accessible to the operator, i.e., the physician. Graphene-based nano-antennas have been shown to efficiently operate in the sub-terahertz band (0.1-10 THz) \cite{jornet2013graphene,akyildiz2014terahertz}, which in turn enables short distance communication through human tissue \cite{elayan2017terahertz}. To ensure resource-efficiency, we propose to rely on THz backscattering communication between sensors and anchors. The main advantage of this technology is, that the passive communication partners (i.e., the BNSs) do not need to actively generate radio-frequency (RF) signals but harvest energy from the signals of the active communication partners (i.e., the anchors) instead.

Up until now, there has been very little research on the localization of in-body BNSs. The main contributions of this paper are a novel, realistic system model for localization and tracking of in-body BNSs and a systematical evaluation of the proposed approach. As the localization performance relies on the placement of the anchors and on the THz backscattering communication, we provide two kinds of evaluation results: i) channel model, path loss and capacity of the THz backscattering communication channel in human tissue and ii) accuracy of the IMU based localization in the cardiovascular system. 

The remainder of the paper is organized as follows. Section \ref{relatedwork} introduces the related work in in-body localization, in THz backscattering localization and in localization based on inertial positioning. The building blocks of our proposed system including the BNSs, the anchor nodes and the wireless communication technology are explained in Section \ref{model}. Section \ref{approach} introduces our localization concept and Section \ref{evaluation} discusses the feasibility of the concept via evaluations. We conclude the paper and outline future work in Section \ref{conclusion}.

\section{Related work}
\label{relatedwork}
The problem of autonomously localizing wireless sensor nodes or mobile robots has been well studied in the past decades, resulting in a large variety of localization algorithms for various types of networks~\cite{simonjan2019adhoc,alrajeh2013localization}. Sensor network localization approaches often exploit signal strength measurements or hop counts of messages while mobile robot localization often relies on inertial positioning. In the related work, we specifically focus on the current state-of-the-art in in-body sensor localization and THz backscattering localization. Furthermore, we briefly discuss inertial positioning methods in mobile networks. 

\subsection{In-body Localization}
To the best of our knowledge, as well as according to \cite{lemic2019survey}, there are only very few attempts towards the localization of in-body BNSs operating in the THz band. Furthermore, all of them focus on localizing the sensors themselves rather than detected anomalies, which is realistically difficult to achieve due to their high mobility. 

One approach was introduced by Tran-Dang et al.~\cite{tran2014localization}, who proposed two hop-counting based localization algorithms. The first technique exploits flooding in order to estimate distances between nanonodes from the number of message hops. To reduce the huge message overhead, the second technique proposed to establish clusters first. Only cluster heads communicate with each other to localize certain nodes. The simulation model assumed a $100\;cm^2$ square area with uniformly distributed nodes and communication ranges of $1-2\;cm$ providing thus very limited evaluation results for the in-body scenario. 

Another method proposed a pulse-based distance accumulation (PBDA) localization algorithm for nanosensor networks which is used to estimate the distance between anchors and clustered nodes \cite{zhou2017pulse}. The simulated system model considered a 2D square area with randomly deployed nodes. First, nodes run a classification algorithm based on flooding which classifies all nodes into corner, border or center nodes. Afterwards, a clustering algorithm establishes clusters and identifies cluster heads. Finally, flooding packets are forwarded through cluster heads in order to calculate the distance between corner nodes based on hops. The algorithm adopted femtosecond-long pulse for terahertz band communication based on on-off keying (OOK) modulation. Similar to \cite{tran2014localization}, the considered simulation model can barely evaluate an in-body scenario.

Shree et al.~\cite{prasad2018direction} introduced an approach in which they utilize the MUltiple SIgnal Classification (MUSIC)~\cite{stoica1989music} algorithm in order to systematically study the Direction of Arrival (DOA) estimation for nanosensor networks. The study was done for different energy levels, distances, pulse shapes and frequencies. In the simulation studies, the authors considered the terahertz channel as standard air medium. Their investigations showed that the error can be reduced by selecting lower order and higher frequency pulses for transmission. A successful estimation of the angle of arrival could enable localization in nanosensor networks, however, sensors would need dedicated antenna arrays. Furthermore, making reliable DOA estimation is difficult in in-body scenarios where sensors are highly mobile, floating inside vessels while frequently changing their orientations. 

In summary, none of the above mentioned localization approaches can be used to provide accurate localization results for the in-body environment. Furthermore, most of the simulations are rather general and do not consider the very specific energy and communication constraints of BNSs. 

\subsection{THz Backscattering Localization}
One promising direction towards obtaining accurate distance measurements, while at the same time maintaining low energy consumption, is to exploit backscattered signals~\cite{lemic2019survey}. Backscattering communication systems consist of two types of devices, a reader (i.e., an anchor) and a tag (i.e., a mobile node)~\cite{liu2019next}. The tag is the passive component of the system, harvesting energy from an incident RF wave radiated by the reader. Further, the tag modulates and reflects a fraction of the wave back to the reader~\cite{liu2019next}. Since tags harvest energy from incoming waves, they do not need active RF components while readers are equipped with such. A well known example for a technology which relies on backscattered signals is radio frequency identification (RFID). 
In order to modulate the received signal, tags change the antenna reflection properties by varying the impedance of the antenna according to the information they want to communicate. As the antenna load is varied, the reflection coefficients are changed and the reflected signal is modulated~\cite{khaledian2019active}. This process is known as backscatter modulation~\cite{dardari2011ultra} and can achieve different modulation modes including binary pulse amplitude modulation (2-PAM), binary pulse position modulation (2-PPM) and On-Off keying modulation (OOK)~\cite{dardari2015method}.
As nanoscale transceivers can not generate carrier signals, they need to exploit carrier-less pulse-based communication schemes in any case~\cite{jornet2014femtosecond,prasad2018direction}. Up until now, there are several works which discuss pulse-based modulations schemes suitable for nanosensor networks, from which Time Spread On-Off Keying (TS-OOK) is the most frequently used~\cite{jornet2014femtosecond, vavouris2018energy}. 

With their low-power and low heat-radiation wireless communication, backscattering tags perfectly suit the integration into BNSs to operate in in-body applications. So far, their feasibility and promising accuracy has only been demonstrated for localization problems on the macro-level \cite{el2018high,lemictoward}. El-Absi et al.~\cite{el2018high} exploited backscattered THz signals to extract the round-trip time-of-flight (RToF) between nodes and anchors to determine the distance between them. Multiple RToF readings from different anchors were used to estimate the location of the node via the linear least square algorithm. There are two main advantages of backscattering. First, the nodes do not need to actively generate RF signals, but only to modulate and reflect received ones. Second, there is no clock drift which could affect the distance estimates since the RToF is determined solely by the reader which is equipped with a high-accuracy clock~\cite{lemic2019survey}. However, the network in \cite{el2018high} was of static nature and thus does not fit an in-body scenario. The second approach \cite{lemictoward} attempted to localize software-defined metamaterial (SDM) elements by utilizing THz backscattering communication for frequencies of $300$ GHz to $10$ THz. The localization relied on trilateration by exploiting distance estimates from RToF measurements between nanonodes and controllers. The authors demonstrated via simulations a sub-millimeter accuracy of the localization as well as a high availability and low energy consumption of the approach. 

Although many next-generation networks are expected to exploit THz communication to provide accurate locations \cite{sarieddeen2020next}, research on localization algorithms for THz-based BNSs in the in-body environment is still lacking.

\subsection{Localization based on Inertial Positioning}
\label{inertialpositioning}
Inertial positioning (or dead reckoning) refers to the process of estimating a node's current position by using its previous position and an estimate about traveled distance and direction which are recorded by a built-in IMU. IMUs typically utilize a combination of accelerometers and gyroscopes to measure acceleration and rotation of the device they are integrated in. Doing so, they are able to determine the angular rate and the linear velocity, and thus the position of the device relative to a global reference system. The major advantage of inertial positioning is that it does not rely on external information enabling a fully autonomous navigation.  

To achieve nanoscale sizes, all BNS components should be able to operate at nanoscale. A variety of nanoscale gyroscopes and accelerometers have already been developed in recent years~\cite{tanner2007feasibility,ahmadian2018novel}. One example of a nanoscale gyroscope was presented by Song et al.~\cite{song2018nanoscale}. The gyroscope has an atom-scale size and can provide a sensitivity in the level of $28$ rad s$^{-1}$ Hz$^{-1/2}$. Besides gyroscopes, graphene-based accelerometers have been presented as well. One example is the accelerometer proposed by Shi et al.~\cite{shi2018modeling} which features a size in the range of tens of $\mu$m and a ultra high sensitivity of $8.82 \cdot 10^{-11}$m/m$\cdot s^{-2}$ for a range of $0-1000$g. In general, inertial positioning is very accurate over short periods with high update frequency. However, the accuracy is significantly affected by the accumulation of noise and drift errors from
accelerometers and gyroscopes. Thus, IMU measurements are often fused with other positioning methods \cite{marquez2017accurate}. 

In the field of robotics, a variety of IMU-based localization systems have been developed to localize and track mobile robots during their missions~\cite{marquez2017accurate,li2019deep,malyavej2013indoor}. 
Li et al. \cite{li2019deep} presented for instance  a deep learning approach to localize a mobile robot using a 2D laser and an IMU. For that purpose, a novel Recurrent Convolutional Neural Network (RCNN) architecture was exploited to fuse the laser and IMU data for pose estimation. 
Another method aiming at accurately locating mobile robots through sensor fusion was proposed by Marquez et al. \cite{marquez2017accurate}. The approach fuses acceleration data from the IMU and 2-D coordinates received from the Ultra-Wideband (UWB) anchors into a Kalman filter to estimate accurate locations. 
A further method for fusion-based localization was introduced by Malyavej et al. \cite{malyavej2013indoor}. The authors proposed to localize mobile robots by exploiting sensor fusion of Received Signal Strength Indicator (RSSI) measurements from a WiFi network and IMU measurements. The proposed fusion is achieved via an extended Kalman filter (EKF).

In general, inertial positioning approaches are expected to be well suited for in-body BNS applications since they operate in a fully autonomous manner and do not require any external information. Naturally, built-in IMUs will add physical complexity to the architecture of BNSs. However, communication and computing complexity can be reduced by this type of self-localization, which eases the requirements on memory and processing unit.

\section{System model building blocks}
\label{model}

\begin{figure}
	\centering
	\def\svgwidth{0.4\textwidth}
\begingroup%
  \makeatletter%
  \providecommand\color[2][]{%
    \errmessage{(Inkscape) Color is used for the text in Inkscape, but the package 'color.sty' is not loaded}%
    \renewcommand\color[2][]{}%
  }%
  \providecommand\transparent[1]{%
    \errmessage{(Inkscape) Transparency is used (non-zero) for the text in Inkscape, but the package 'transparent.sty' is not loaded}%
    \renewcommand\transparent[1]{}%
  }%
  \providecommand\rotatebox[2]{#2}%
  \newcommand*\fsize{\dimexpr\f@size pt\relax}%
  \newcommand*\lineheight[1]{\fontsize{\fsize}{#1\fsize}\selectfont}%
  \ifx\svgwidth\undefined%
    \setlength{\unitlength}{765.35433071bp}%
    \ifx\svgscale\undefined%
      \relax%
    \else%
      \setlength{\unitlength}{\unitlength * \real{\svgscale}}%
    \fi%
  \else%
    \setlength{\unitlength}{\svgwidth}%
  \fi%
  \global\let\svgwidth\undefined%
  \global\let\svgscale\undefined%
  \makeatother%
  \begin{picture}(1,0.47407407)%
    \lineheight{1}%
    \setlength\tabcolsep{0pt}%
    \put(0,0){\includegraphics[width=\unitlength,page=1]{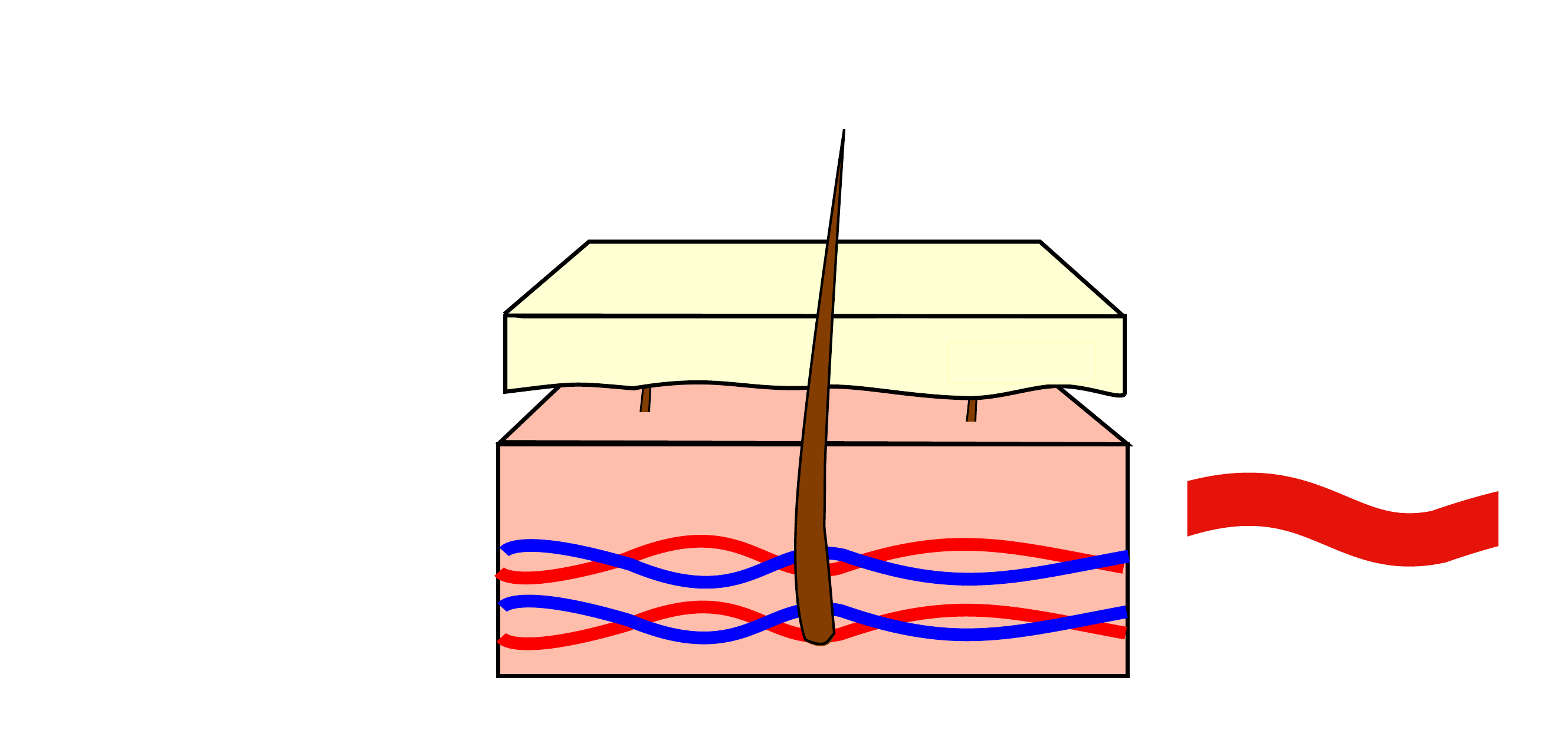}}%
    \put(0.03764765,0.25301975){\color[rgb]{0,0,0}\makebox(0,0)[lt]{\lineheight{0}\smash{\begin{tabular}[t]{l}Epidermis\end{tabular}}}}%
    \put(0,0){\includegraphics[width=\unitlength,page=2]{SkinLayers.pdf}}%
    \put(0.74249258,0.05718246){\color[rgb]{0,0,0}\makebox(0,0)[lt]{\lineheight{0}\smash{\begin{tabular}[t]{l}Nanosensor\end{tabular}}}}%
    \put(0,0){\includegraphics[width=\unitlength,page=3]{SkinLayers.pdf}}%
    \put(0.41190353,0.43600425){\color[rgb]{0,0,0}\makebox(0,0)[lt]{\lineheight{1.25}\smash{\begin{tabular}[t]{l}Anchor node\end{tabular}}}}%
    \put(0.03610314,0.13240999){\color[rgb]{0,0,0}\makebox(0,0)[lt]{\lineheight{1.25}\smash{\begin{tabular}[t]{l}Dermis with\\blood vessels\end{tabular}}}}%
  \end{picture}%
\endgroup%

	\caption{Simplified skin layer model, including the epidermis and the dermis with the blood vessels. BNSs float through the vessels while anchors are attached to the skin.}
	\label{fig:SkinLayers}
\end{figure}

Our proposed system comprises of two kinds of devices, the BNSs which float through the cardiovascular system and the anchor nodes which are attached to the skin. The BNSs are tiny sensing devices built with nanomaterials (e.g., graphene) which are injected into the human bloodstream in order to monitor certain health parameters~\cite{khan2020nanosensor,santiago2018nanoscale}. The anchors are larger in scale, more powerful, and they are intended to collect sensing and location information from the sensors as they float by. 
Figure \ref{fig:SkinLayers} depicts an overview of our system model. We assume a simplified model of the skin layers including the epidermis and the dermis which in turn contains the blood vessels with the floating sensors. 

\subsection{Anchor Nodes}
Anchors are attached to the skin across the whole body of the human who is being monitored. 
The anchors are the active components of the backscattering communication system, and serve as basestations and gateways to enable access to the measurements of the BNSs. Anchors constantly transmit RF signals which are used by the sensors for the backscattering communication. Whenever a sensor is sufficiently close to an anchor and receives its RF signal, it harvests energy from the signal, modulates it to pack its data into the signal and reflects it back to the anchor. 

Communication among anchors or between anchors and a sink node (i.e., computer) can be achieved via a standard networking technology such as IEEE $802.11$ or $802.15$. Our localization concept foresees anchors to maintain tables in which they collect all the received information of the sensors. These tables are periodically sent to a sink computer, where they are merged in order to keep an updated view of the sensors and their measurements. Anchors are synchronized in time to be able to timestamp received packets, while sensors do not need any synchronization.

\subsection{Bionanosensors}
Bionanosensors, i.e., BNSs, are nanoscale devices which are able to detect, (pre-)process and transfer sensed data. Thus, they comprise sensing, communication and processing unit. Depending on the physical parameter to be monitored, the sensing unit requires a certain type of sensor, e.g., pressure sensor~\cite{li2018ultrahigh}, temperature sensor~\cite{rajasekar2019nano}, mechanical sensor~\cite{qu2019eccentric} or optical sensor~\cite{simonjan2018nano}. Since the BNSs should provide sensor readings along with location information, they are further equipped with IMUs to measure velocity and rotation. 
Incorporating IMUs into the BNSs will allow for inertial positioning methods similar to those used in mobile robot applications (see Section \ref{inertialpositioning}). As already mentioned, communication between BNSs and anchors relies on backscattering with the sensors being the passive communication partners. Thus, sensors do not need to be equipped with an active RF component. However, they need an energy harvester, a receiver and a modulator in order to exploit and modulate the received signal before reflecting it back to the anchor. Similar to standard sensors, BNSs also require a memory and a battery unit to be able to operate their local sensors. 

\subsection{Wireless Communication}
Currently, the material most commonly associated with nanoantennas is graphene, which is a flat monolayer of carbon atoms tightly packed into a 2D honey-comb lattice~\cite{jornet2013graphene}. Graphene antennas efficiently operate at Terahertz band frequencies (0.1-10 THz)~\cite{jornet2013graphene,jornet2014femtosecond}. With their small size of just tens of nanometers in width and a few micrometers in length, they can easily be integrated into future BNSs. 

The maximum distance for successful communication between a sender and a receiver depends on the characteristics of the THz channel. The main issue in THz band communication is the very high path loss due to the molecular absorption loss~\cite{KOKKONIEMI201635,jornet2011channel}. Molecular absorption loss is caused by the water molecules in the medium that are excited by the electromagnetic wave, which leads to the conversion of a part of the electromagnetic wave energy into kinetic energy. 
However, there are several works which investigated the path loss of THz signals in blood and tissue, showing that a communication in the range of a few millimeters is feasible \cite{elayan2017terahertz,piro2016terahertz}. For the in-body scenario considered here, communication distances of a few millimeters are sufficient to establish connection between the anchors and BNSs.

\section{Localization approach}
\label{approach}
\begin{figure}
	\centering
	\def\svgwidth{0.42\textwidth}
		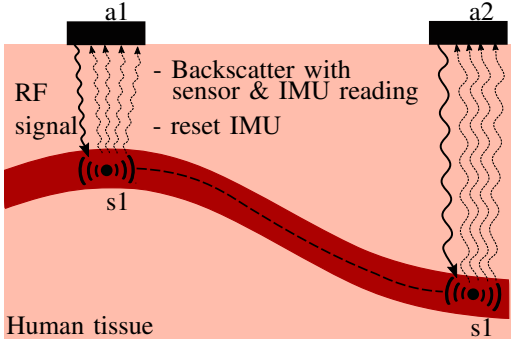
	\caption{Overview of the proposed in-body localization. Whenever BNS $s1$ is sufficiently close to an anchor ($a1$ or $a2$), it exploits the RF signal to transmit its sensor readings. After each transmission, the sensor resets its IMU to the initial state to decrease the accumulation error.}
	\label{fig:Localization}
\end{figure}

As BNSs are envisioned to float with average velocities of $10-20$ cm/s in the bloodstream \cite{fruchard2013estimation}, keeping an updated view of their locations at any point in time is complex in terms of computation and communication costs. 
Rather than continuously tracking the highly dynamic sensors, we propose to record the location information (IMU readings) along the sensor measurements every time the sensor encounters an anomaly. This means, whenever the sensors relay their measurements to anchors, the locations of the anomalies will be transferred as well. This way, we are able to determine the location of the anomaly in the body. 
As sensors can exploit backscattering only within the vicinity of an anchor, anchors implicitly gather information about their locations via receiving their messages. Figure \ref{fig:Localization} depicts an overview of the localization system. Sensor $s1$ is floating inside the blood vessels while the anchor nodes $a1$ and $a2$ are attached to the skin. 
Whenever the sensor passes by an anchor, it exploits the RF signal to transmit collected sensor readings along with the IMU measurements via backscattering communication. Each packet transmitted by a BNS includes the ID of the sensor and the actual sensor reading with a location stamp. Whenever a packet was transmitted to an anchor, the sensor resets its IMU to the initial state. The reset is done in order to reduce the accumulation error (i.e., the drift) of the IMU. The following subsections discuss the backscattering communication and the location stamping in detail.

\subsection{Backscattering Communication}
As already mentioned, a promising technique for low-power wireless communication is backscattering communication. In order to prove the feasibility of backscattering communication for in-body scenarios, we model the backscattering channel between anchors and BNSs and determine its capacity. The channel consists of a forward path (reader to tag) and a backward path (tag to reader). To be able to calculate the capacity of the complete channel, forward and backward link need to be taken into account. The backscattered power in dB received at the reader after the signal travelled through the forward and backward link can be expressed as \cite{griffin2009complete} 

\begin{equation}
 P_{R_B} [dB] = P_T  + G_T - 2L_{tot} + G_R,
\label{recpower}
\end{equation}
where $P_T$ is the transmit power, $L_{tot}$ is the total path loss of the signal and $G_T$ and $G_R$ are the gains of the transmitting and receiving antennas, respectively. The total path loss $L_{tot}$ of the signal mainly depends on the spreading and the molecular absorption loss as the scattering loss can be neglected \cite{elayan2017multi} and it is defined as 

\begin{equation}
 L_{tot}(f) = L_{abs}(f) \times L_{spr}(f) = e^{-\mu_{abs} d} \times \left(\frac{\lambda_g}{4\pi d}\right)^2,
\label{pathloss}
\end{equation}
where $L_{abs}$ is the molecular absorption loss, $L_{spr}$ is the spreading loss, $\mu_{abs}$ is the molecular absorption coefficient, $d$ is the distance and $\lambda_g$ is the effective wavelength, which is defined as $\lambda/n'$. $n'$ and $n''$ are the real and imaginary parts of the refractive index $n$ of the medium the wave is travelling through (tissue in our case). The molecular absorption coefficient $\mu_{abs}$ depends on the effective wavelength and on the imaginary part of the refractive index and is defined as \cite{elayan2017terahertz} 

\begin{equation}
    \mu_{abs} = \frac{4 \pi n''}{\lambda_g}.
    \label{muabs}
\end{equation}

As backscattering reduces the power significantly, also channel capacity is affected. The channel capacity $C_B$ of the forward and the backscattered communication link  can be expressed using the Shannon theorem and a flat power distribution as \cite{jornet2011channel, piro2016terahertz}
\begin{equation}
	C_B = \sum \Delta f \cdot log_2 \left(1+ \frac{S(f)}{L_{tot}(f,d)N(f,d)}\right),
	\label{capacity}
\end{equation}
where $S(f)$ is the power spectral density of the signal, $\Delta f$ is sub-band width, $L_{tot}(f,d)$ is the total loss over the stacked tissue layers (obtained by Equation \ref{pathloss}) and $N(f,d)$ is the noise power spectral density. If the sub-band width is small enough, the
channel appears as frequency-non selective and the signal power spectral density $S(f)$ can be considered locally flat \cite{jornet2011channel}. The noise spectral density can then be obtained as \cite{yang2015numerical, piro2016terahertz} 
\begin{equation}
	N(f,d) = k_B \cdot T_{mol}
	= k_B \cdot T_0 \left(1-e^{-4\pi fdn'' / c}\right),
\label{eq:noise}
\end{equation}					

where $k_B$ is the Boltzmann constant, $T_{mol}$ is the noise temperature due to molecular absorption, $T_0$ is the reference temperature of $310$ K, $f$ is the frequency, $d$ is the distance, $c$ is the speed of light and $n''$ is the imaginary part of the refractive index of the respective tissue.


\subsection{Location Stamping}
\label{locationstamping}
To be able to provide sensor readings with location stamps, BNSs rely on inertial positioning (see Section \ref{inertialpositioning}). 
We propose to equip BNSs with IMUs including a nanoscale gyroscope and accelerometer to be able to measure rotation and velocity and thus get an idea of how far they have traveled within the body. 
As already discussed, the major drawback of inertial positioning is the accumulation of the error as it is a recursive process relying on faulty sensor readings. Thus, positions need to be refined (i.e., through sensor fusion) or the IMU needs to be reset whenever possible in order to minimize the error. Therefore, we propose that BNSs reset their IMU to the initial state whenever they have transmitted their readings to an anchor. 
Having a BNS communicating to an anchor infers that the sensor is in very close proximity to the anchor, which implicitly helps the sensor to understand its actual location enabling to refine the position calculated by the IMU. The communication between sensors and anchors allows the system thus to additionally keep track of the sensors' positions in the body. 

The true location of an anchor is $(x_a,y_a,z_a)$ and the location logged by the BNS passing by this anchor is $(\hat{x}_a,\hat{y}_a,\hat{z}_a)$, where $\hat{x}_a=x_a + \epsilon_x$, $\hat{y}_a=y_a + \epsilon_y$, $\hat{z}_a=z_a + \epsilon_z$, with $\epsilon_x$, $\epsilon_y$ and $\epsilon_z$ being uniformly distributed errors within the bounds determined by the range of backscattering communication. 
Upon visiting an anchor and resetting its location, the BNS starts calculating its location using the built-in IMU. However, IMUs, especially the novel nanoscale IMUs, introduce measurement errors in all directions which are expressed for the accelerometer as
\begin{equation}
    \Tilde{a}(t) = a(t) + b_a + \eta_a,
\end{equation}
where $b_a$ is the bias and $\eta_a$ is random Gaussian noise. Similarly, the gyroscope introduces measurement errors in all three directions expressed as %
\begin{equation}
    \Tilde{\omega}(t)=\omega(t) + b_g + \eta_g
\end{equation}
where $b_g$ is the bias and $\eta_g$ is random Gaussian noise. 

To calculate their current positions, BNSs combine accelerometer and gyroscope readings. Afterwards they utilize a Kalman filter to calculate their current position based on the previous one acquired at the last anchor node \cite{alatise2017pose}. 
When a BNS senses an anomaly along its path, it records the location of the event to be able to report it to the next anchor.   

Similar to anchor locations, the actual location of an anomaly is expressed $(x_e,y_e,z_e)$. The anomaly location estimated by the BNS is expressed as $(\hat{x}_e,\hat{y}_e,\hat{z}_e)$, including the errors mentioned above. 


\section{Evaluation}
\label{evaluation}

This section presents the evaluation results for our proposed localization approach. We first discuss the channel characteristics of in-body THz backscattering communication, followed by a discussion on sensor movement and localization accuracy. 

\subsection{Communication}
\label{communicationEval}
This section investigates the feasibility of THz backscattering communication between BNSs in the blood and anchors attached to the skin. To enable a realistic channel model, certain biological parameters such as human skin thickness and blood velocity are used in evaluation. The volar forearm, the scalp, the dorsums of the hand and the anterolateral thigh have been shown to have average epidermal thicknesses of $90\; \mu$m, $108\; \mu$m, $181\; \mu$m and $60\; \mu$m, respectively \cite{oltulu2018measurement,chan2014skin}. The average dermal thicknesses for the aforementioned regions have been measured to be $1100\; \mu$m, $1984\; \mu$m, $1864\; \mu$m and $1950\; \mu$m \cite{oltulu2018measurement,akkus2012evaluation}. Based on these measurements, we used simulation values of $200\; \mu$m and $1800\; \mu$m for epidermal and dermal thickness, respectively. Blood vessels vary a lot in terms of their diameter with $10^{^-5}\;$m being the smallest in the capillaries, followed by $10^{-4}\;$m in the arterioles and $10^{-3}\;$m in the small arteries \cite{fruchard2013estimation}. For our simulation studies, we thus picked an average value of $500\; \mu$m for the thickness of the blood layer. The chosen values add up to a total simulated tissue thickness of $2500\; \mu$m.  

In general, the propagation of THz-band waves in tissue is drastically impacted by the absorption of liquid water molecules which cause internal vibrations (i.e., absorption loss). The absorption loss is determined using the absorption coefficient of the respective tissue layer. To be able to calculate the absorption coefficient using Equation \eqref{muabs}, the dielectric characteristics of the tissue layers are required. Table \ref{tab:permittivity} summarizes the dielectric parameters of epidermis, dermis and blood which we used in our simulations. The complex permittivity $\varepsilon_r$ of tissues with a high amount of water molecules (i.e., blood) can be approximated best by the double Debye relaxation \cite{elayan2017multi}:

\begin{equation}
    \varepsilon_r = \varepsilon_\infty + \frac{\varepsilon_1-\varepsilon_2}{1 + j\omega \tau_1} + \frac{\varepsilon_2 - \varepsilon_\infty}{1 + j\omega \tau_2},
\end{equation}

where $\omega$ is the angular frequency, $\varepsilon_\infty$ is the relative permittivity at the high frequency limit, $\varepsilon_1$ and $\varepsilon_2$ are the amplitude changes and  $\tau_1$ and $\tau_2$ are the characteristic relaxation times of the medium. Compared to that, the complex permittivity $\varepsilon_r$ of skin and fat tissue can be modelled best by the Havriliak-Negami relationship~\cite{piro2016terahertz}:    

\begin{equation}
    \varepsilon_r = \varepsilon_\infty + \sum_{i=1}^{N} \frac{\varepsilon_i}{[1 + (j\omega \tau_i)^{\alpha_i}]^{\beta_i}} - j \frac{\sigma}{\omega \varepsilon_{0}},
\end{equation}

where the additional exponents $\alpha$ and $\beta$ describe the asymmetry and broadness of the corresponding spectra and $\sigma$ the static ionic conductivity.

\begin{table}
\begin{center}
\resizebox{0.38\textwidth}{!}{
\begin{tabular}{l c c c}
\rowcolor{Gray} 
\hline \textbf{Parameter} & \textbf{Blood} & \textbf{Dermis} & \textbf{Epidermis} \\ \hline 
$\alpha_1$ 						& 				& $0.92$			& $0.95$ 		\\ 
\rowcolor{Lightgray}
$\alpha_2$ 						& 				& $0.97$			& 					\\ 
$\beta_1$ 						& 				& $0.8$				& $0.96$		\\ 
\rowcolor{Lightgray}
$\beta_2$ 						& 				& $0.99$			&						\\ 
$\varepsilon_\infty$ 	& $2.1$		& $4$					&  $3.0$		\\ 
\rowcolor{Lightgray}
$\varepsilon_1$ 			& $130$		& $5.96$			&  $89.61$	\\ 
$\varepsilon_2$ 			& $3.8$		& $380.4$			&  					\\ 
\rowcolor{Lightgray}
$\tau_1$ (ps) 				& $14.4$	& $1.6$				&  $15.9$		\\ 
$\tau_2$ (ps) 				& $0.1$ 	& $159$ (ns)	& 					\\ 
\rowcolor{Lightgray}
$\sigma$ 							& 				& $0.1$ 			&  					\\ 
Ref. & \cite{reid2013terahertz} & \cite{piro2016terahertz} & \cite{piro2016terahertz} \\ \hline
\end{tabular}}
\caption{Permittivity and relaxation time values for blood, dermis and epidermis.}
\label{tab:permittivity}
\end{center}
\end{table}

\begin{figure}
	\centering
		\includegraphics[width=0.375\textwidth]{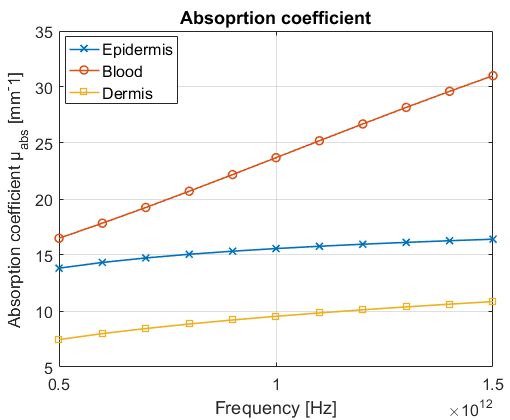}
	\caption{Absorption coefficient $\mu_{abs}$ of epidermis, dermis and blood in the sub-terahertz band.}
	\label{fig:absCoeff}
\end{figure}

 Figure \ref{fig:absCoeff} shows the absorption coefficient of the different tissue layers at sub-THz frequencies. As we can see from the figure, the absorption loss is more dominant in blood than in the other tissue layers, which is due to the fact that blood contains the highest amount of water molecules.  

\begin{figure}
\centering
\begin{subfigure}{.38\textwidth}
  \centering
  \includegraphics[width=\linewidth]{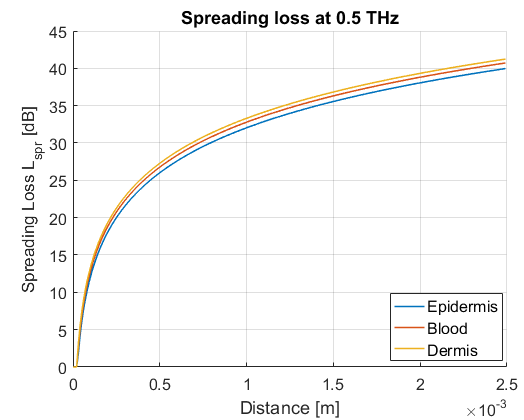}
		\caption{}
\end{subfigure}
\begin{subfigure}{.38\textwidth}
  \centering
  \includegraphics[width=\linewidth]{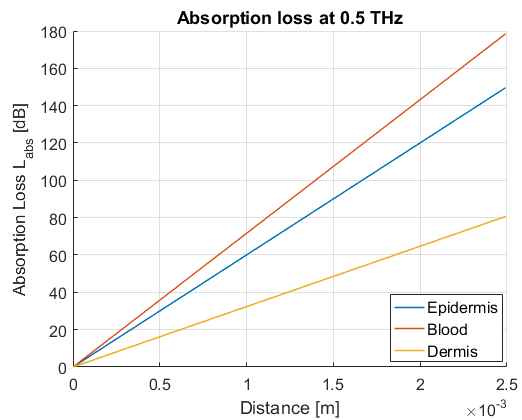} 
	\caption{}
\end{subfigure}
\caption{Spreading loss (a) and absorption loss (b) at $0.5$ THz for the three tissue layers over distance.}
\label{fig:losses}
\end{figure}

Using the absorption coefficient $\mu_{abs}$ and the equations for spreading and absorption loss (see \eqref{pathloss}), we determined the spreading and the absorption loss for the three tissue layers. Figure \ref{fig:losses} shows the results for the spreading loss (a) and the absorption loss (b) in dB at $0.5$ THz. As can be seen from the figures, the signal suffers significantly more from molecular absorption than from spreading loss. 

\begin{figure}
	\centering
		\includegraphics[width=0.38\textwidth]{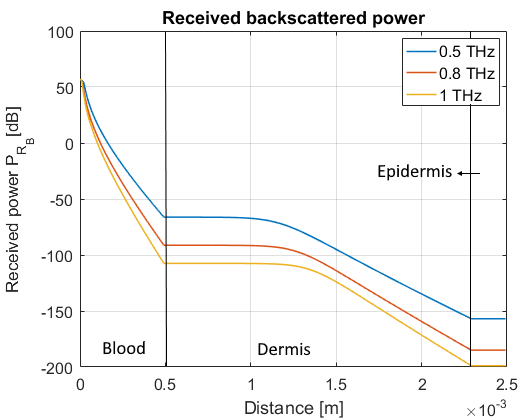}
	\caption{Received backscattered power $P_{R_{B}}$ in dB for frequencies of $0.5, 0.8$ and $1$ THz after travelling through the three stacked tissue layers twice. }
	\label{fig:backscatter}
\end{figure}

Since we propose to rely on backscattering communication, we are specifically interested in the backscattered power received at the anchor and defined by \eqref{recpower}. Therefore, we stacked the three tissue layers and modeled a signal traveling the channel twice (from the anchor to the BNS and back). The thicknesses of the layers were thereby chosen as discussed above and the antenna gains $G_T$ and $G_R$ were chosen to be $5.09$ \cite{seyedsharbaty2017antenna}. In line with \cite{piro2016terahertz,jornet2011channel}, the pulse energy and the pulse duration are assumed to be constant at $500$ pJ and $100$ fs, resulting in a transmitted peak power of $5$ kW. 
Figure \ref{fig:backscatter} shows the received backscattered power $P_{R_{B}}$ in dB for frequencies of $0.5, 0.8$ and $1$ THz when being transmitted through $500\; \mu$m of blood, $1800\; \mu$m of dermis and $200\; \mu$m of epidermis (adding up to a total distance of $2500\; \mu$m). As can be seen from the figure, after traveling twice through the tissue, the signal power received at the anchor is approx. $-156$, $-185$ and $-198$ dB depending on the used frequency. 

Receiving a signal at $-156$ dB requires the receiver to have a sensitivity of approx. $0.25$ fW/Hz$^{1/2}$. The sensitivity requirement can be eased by applying a higher transmit power or by using higher gain antennas. Nonetheless, nanoelectronic detectors with sensitivities of $10^{-15}$ - $10^{-20}$ W/Hz$^{1/2}$ have already been proposed in literature \cite{rogalski2019two}. 

The state-of-the-art further shows that channel capacities on the order of Gbps-Tbps can be achieved for distances below $4\;$ mm in in-body THz communication scenarios \cite{piro2016terahertz}. As backscattering reduces the power significantly, channel capacity is also affected. We calculated the channel capacity $C_B$ of the forward and the backscattered communication link using \eqref{capacity}. We used the initial transmission power of the anchor to determine $S(f)$ for the forward link, and the power received at the BNS to determine $S(f)$ for the backscattered link. 

\begin{figure}
	\centering
		\includegraphics[width=0.365\textwidth]{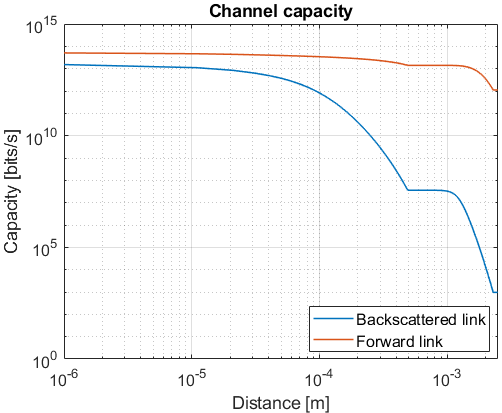}
	\caption{Channel capacity of the forward and the backscattered link for the stacked tissue layers (total distance of $2500\; \mu$m).}
	\label{fig:capacity}
\end{figure}

Figure \ref{fig:capacity} shows the theoretical capacity of the forward and the backscattered communication link of the stacked tissue layers. In line with \cite{piro2016terahertz}, the capacity for the forward link at a distance of $2500\; \mu$m is in the order of Tbps. However, the capacity of the backscattered link is drastically decreased to approx. $1$ kbps. To achieve higher capacities in backscattering communication, the system will require significantly higher transmit power and/or antenna gains. Generally speaking, channel capacities on the order of kbps are more realistically achievable than Gbps. 

Summarized, sub-THz backscattering communication in human tissue is feasible for very limited distances. Specifically, the anchors should be attached to regions of the body where the skin and fat layers are thin, i.e., volar forearm, scalp or dorsum of the hand. 

\subsection{Localization}
This section studies the movement of the BNSs in the bloodstream, the requirements on the anchor placement and the localization accuracy of our concept.  
There are two factors which mainly influence the placement of the anchors. First, THz backscattering enables communication in tissue for distances of $2500 \; \mu $m at most. Anchors need thus to be placed at body regions with thin skin and fat layers. The second requirement on the placement of anchors comes from the errors generated by the inertial positioning process. As discussed in Section \ref{locationstamping}, the accelerometer and the gyroscope in the IMU suffer from drift errors inducing accumulation errors over time. Since BNSs reset their IMUs whenever they communicate with an anchor, anchors should not be placed too far from each other in order to keep accumulation errors small. The following two subsections discuss the exploited simulation environment and the localization results.

\paragraph{Simulation Environment}
Modeling all characteristics of the cardiovascular system is non-trivial as it comprises approx. $4900$ cm$^3$ of blood volume and $120\;000$ km of blood vessels \cite{tortora2018principles}. The blood flow rate depends on the diameter of the vessel reaching from $10-20$ cm/s in the arteries over $0.1$ cm/s in the arterioles to $5 \cdot 10^{-3}$ cm/s in the capillaries \cite{fruchard2013estimation}. 

There is no simulation tool available yet, which incorporates the entire circulatory system in detail. However, to achieve realistic estimates about our localization concept, a simplified model including all major vessels of the cardiovascular system is sufficient. Therefore, we used BloodVoyagerS \cite{geyer2018bloodvoyagers}, which is a medical nanonetwork simulation module for ns-3\footnote{https://www.nsnam.org/}. BloodVoyagerS is modeled upon a simplified human cardiovascular system to realize the movement of nanobots within the human bloodstream. The simulator comprises a model including $94$ vessels and their respective blood flow rates ($20$ cm/s in the aorta, $10$ cm/s in the arteries, $2-4$ cm/s in the veins), adding up to a total simulated vessel length of $12\;717$ m.

\begin{figure}
	\centering
		\includegraphics[width=0.45\textwidth]{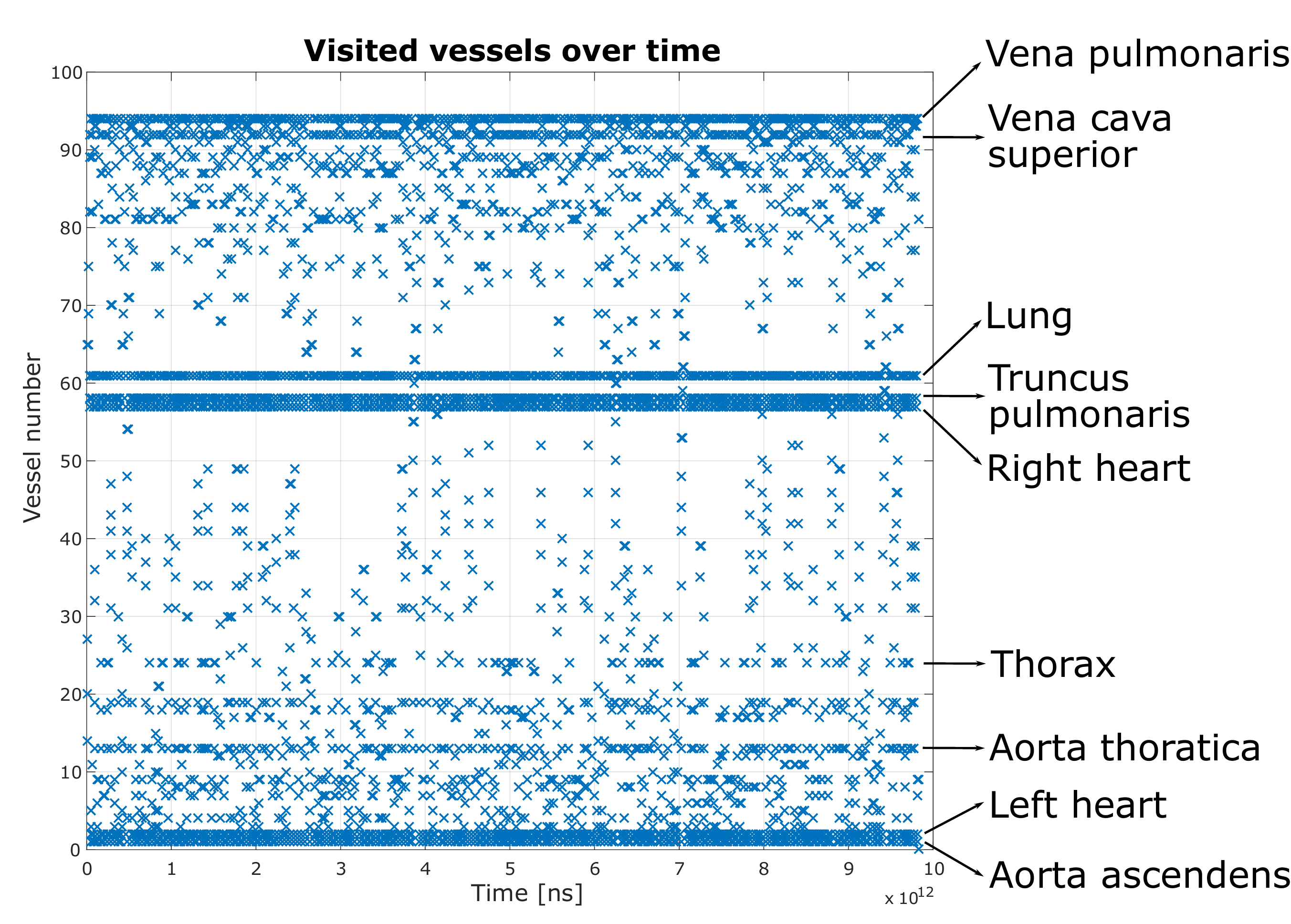}
		\caption{Visited vessels over time.}
	\label{vessels}
\end{figure}

\begin{figure}
	\centering
		\includegraphics[width=0.35\textwidth]{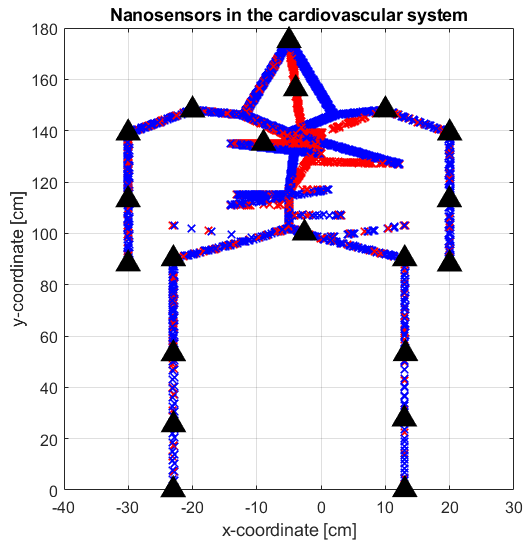}
		\caption{Simulation results of one BNS moving through the cardiovascular system for $10\;000$ s with a setup of $20$ anchors. Red and blue crosses depict the artery and vein locations of the BNS, respectively. The black triangles depict the anchor locations.}
	\label{bodies}
\end{figure}

\paragraph{BNS Movement}
In a first step, we were interested in the movement of the BNSs, in order to get an idea which parts of the body are visited the most. For that purpose, we performed a simulation of one BNS moving in the cardiovascular system for $10\;000$ s. Figure \ref{vessels} shows which vessels have been visited by the BNS at each time step. In BloodVoyagerS, all vessels and organs have dedicated numbers. The y-axis of Figure \ref{vessels} shows the numbers of the respective vessel (refer to \cite{geyer2018bloodvoyagers} for all vessel numbers) and the x-axis shows the time step in ns. We annotated the most visited vessels to show which body locations are reached more frequently than others. As expected, the BNS passes more often through upper body regions such as the heart and the lungs than the extremities. This result suggests to place more anchors around the upper body to have BNSs floating by anchors more often.\\ 
In a next step, we deployed $20$ anchors around the body and logged the locations of one BNS to study how often it passes by an anchor. We injected the BNS into the brachial artery (upper left arm) and tracked its movement for $10\;000$ s (approx. $2.7$ h). Figure \ref{bodies} shows the simulation results in terms of BNS locations (x and y) recorded over time and depicted by red (artery) and blue (vein) crosses. The black triangles depict the anchor locations. Using the track of the BNS, we determined each time step at which the sensor was in close proximity to an anchor (i.e., the BNS is within a $2.5$ cm radius of an anchor as we assumed anchors to be in the shape of $5 x 5\;$cm$^2$ patches). We ran the simulation $50$ times and determined the average duration between two anchor visits. Almost $80 \%$ of consecutive anchor visits occurred within a time window of $10$ s. The studies showed that in the worst case, a visit can take up to $80$ s in the $20$ anchor setup. However, considering that $80 \%$ of the time, anchor visits occur within $10$ s, an appropriate sampling rate for acquiring sensor readings could be chosen as $0.1-0.2$ Hz (sensor measurements are performed every $5-10$ s). The sampling rate depends on the requirements of the application, however, sampling rates of up to $4$ Hz are realistically achievable for current BNS prototypes \cite{li2018ultrahigh,qu2019eccentric}.

\paragraph{Event Localization}

Here, we investigate the event localization of the presented system by BNSs reporting to anchors. BNSs stamp the location of an event representing an anomaly whenever they sense one (as described in Section \ref{locationstamping}). Since we assume anchors to be shaped as $5 x 5\;$cm$^2$ patches, sensors communicating to an anchor log their location as random point within this square. Thus, the x and y components of the location can take any value within $[-2.5,2.5]$. Since the skin thickness is assumed to be $2500$ $\mu m$ on average, BNSs infer their z direction from the interval $[-0.25,0]$. The upper bound is $0$ as the anchor can not be outside the body. 
\\
To analyze the performance of the localization system, we used MATLAB's Sensor Fusion and Tracking Toolbox\footnote{https://www.mathworks.com/products/sensor-fusion-and-tracking.html} along with BloodVoyagerS. The ground truth trajectories of the BNSs are determined using BloodVoyagerS and the Kalman filter based location stamping is performed with the MATLAB toolbox. For the evaluation, we used multiple BloodVoyagerS runs, calculated the localization error for each run and averaged the results.

\begin{figure}
        \begin{subfigure}[b]{0.49\columnwidth}
            \centering
            \includegraphics[width=1\columnwidth]{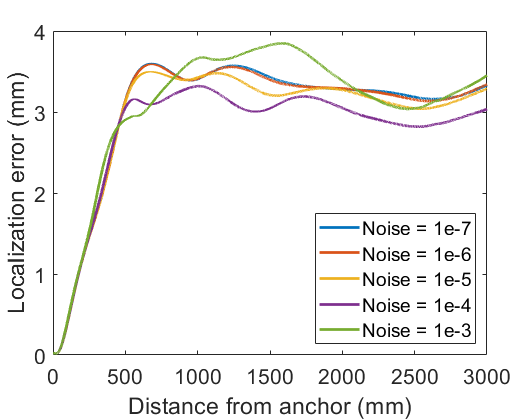}
            \caption[]%
            {{\small }}    
            \label{fig:gnoise}
        \end{subfigure}
        \hfill
        \begin{subfigure}[b]{0.49\columnwidth}  
            \centering 
            \includegraphics[width=1\columnwidth]{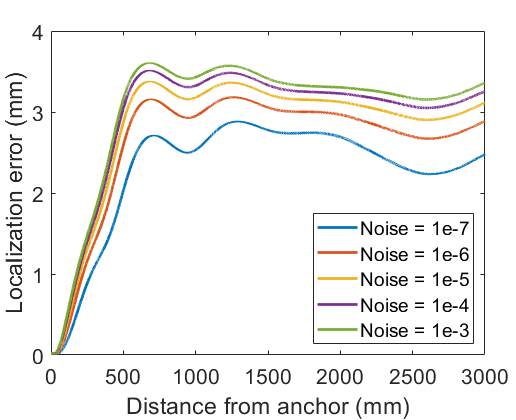}
            \caption[]%
            {{\small}}    
            \label{fig:anoise}
        \end{subfigure}
        \vskip\baselineskip
        \begin{subfigure}[b]{0.49\columnwidth}   
            \centering 
            \includegraphics[width=1\columnwidth]{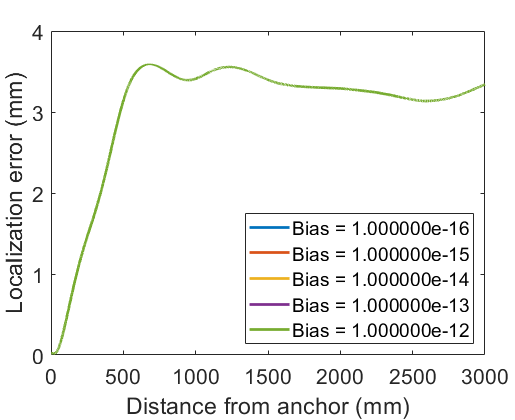}
            \caption[]%
            {{\small }}    
            \label{fig:gbias}
        \end{subfigure}
        \hfill
        \begin{subfigure}[b]{0.49\columnwidth}   
            \centering 
            \includegraphics[width=1\columnwidth]{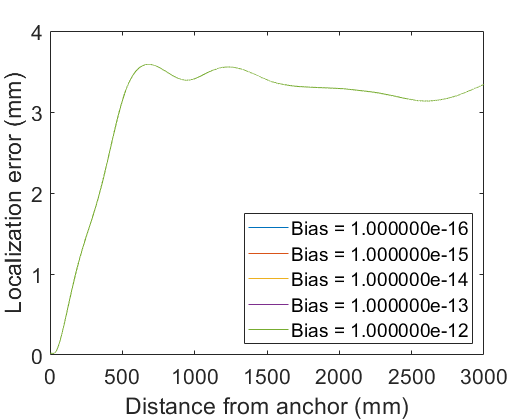}
            \caption[]%
            {{\small }}    
            \label{fig:abias}
        \end{subfigure}
        \caption[  ]
        {\small Localization error for varying a) gyroscope noise b) accelerometer noise c) gyroscope bias d) accelerometer bias} 
        \label{fig:mean and std of nets}
    \end{figure}

We consider a single BNS sensing the event. Figure \ref{fig:mean and std of nets} illustrates the localization error over increasing distance between the BNS and the anchor. We calculated the localization error over varying sensor noise and bias levels in accordance with typical values for nanoscale IMUs. 
Figure \ref{fig:mean and std of nets} (a) and (b) show the localization error with varying noise levels (and fixed bias) for the gyroscope and the accelerometer, respectively. From the figure, we can observe that the localization error first increases with increasing distance to the anchor (up to approx. $500$ mm). This matches our expectations as the IMU error accumulates over time. 
The error then begins to stabilize at a value of approx. $3.2\;$mm which is due to the Kalman Filter.  
Figure \ref{fig:mean and std of nets} (c) and (d) depict the bias of the gyroscope and the accelerometer (with fixed noise), respectively. As can be seen from the figures, all bias levels result in the same line, which is further very similar to the noise level plots. This means that the localization error is not significantly affected by the varying IMU bias. Comparing Figure \ref{fig:mean and std of nets} (a) to \ref{fig:mean and std of nets} (c) and Figure \ref{fig:mean and std of nets} (b) to \ref{fig:mean and std of nets} (d), we can conclude that the contribution of noise is more significant than the contribution of bias.   
Since IMU errors accumulate over time, it is important to update BNS locations at the anchors to bound the localization errors. To keep the localization error below $2\;$mm, anchors should be placed such that the distance between them does not exceed $50\;$cm. However, even a localization error of $3.5\;$mm is very reasonable as we consider monitoring the human body which has an average height of $1.75\;$m.

\section{Conclusion}
\label{conclusion}
Traditional localization approaches are not feasible for in-body scenarios due to the specific communication and resource constraints of nanosensors as well as the highly dynamic environment. This paper has proposed a novel localization concept for bionanosensors floating through the cardiovascular system. The outlined system comprises nanosensors floating through the bloodstream as well as anchors attached to the human skin, and relies on inertial positioning and THz backscattering communication. Simulations have been performed as a proof of concept and provide promising results. As nanosensors are still in the development process, our approach can be seen as a starting point for the development of localization methods for future health monitoring applications. 
To further improve our concept, future work needs to be investigated with an interdisciplinary team to explore human health and safety considerations as well as the feasibility of the proposed nanosensors.

\bibliographystyle{myIEEEtran}
\bibliography{literature}

\end{document}